\documentclass[sigconf]{acmart}
\usepackage{colortbl}
\usepackage{threeparttable}
\usepackage{multirow}
\usepackage{graphicx,pifont}
\usepackage{xcolor}

\usepackage{amsmath,amssymb,amsfonts}
\usepackage{textcomp}
\usepackage{algorithm}
\usepackage{algorithmicx}
\usepackage{algpseudocode}
\usepackage{geometry}
\renewcommand\footnotetextcopyrightpermission[1]{}

\acmConference[ICCAD'2024]{International Conference on Computer-Aided Design}{October 27--31, 2024}{New Jersey, USA}

\begin{document}

\title{SNNGX: Securing Spiking Neural Networks with Genetic XOR Encryption on RRAM-based Neuromorphic Accelerator}

\author{\fontsize{10}{8}\selectfont Kwunhang Wong$^{1,2,*}$, Songqi Wang$^{1,2,*}$, Wei Huang$^{2}$, Xinyuan Zhang$^{1,2}$, Yangu He$^{1,2}$, Karl M.H. Lai$^{2}$, \\ Yuzhong Jiao$^{1}$, Ning Lin$^{1,2,\dag}$, Xiaojuan Qi$^{2}$, Xiaoming Chen$^{3,\dag}$ and Zhongrui Wang$^{1,2,\dag}$}

\affiliation{%
  \institution{\fontsize{8}{8}\selectfont$^1$ACCESS – AI Chip Center for Emerging Smart Systems, InnoHK Centers, Hong Kong Science Park, Hong Kong; $^2$Department of Electrical and Electronic Engineering, The University of Hong Kong, Hong Kong; $^3$Institute of Computing Technology, Chinese Academy of Sciences, Beijing, China; 
  \\ $^*$Equal contribution to this work; $^\dag$Corresponding authors: linning@hku.hk; chenxiaoming@ict.ac.cn; zrwang@eee.hku.hk}
  \country{}
}

\renewcommand{\shortauthors}{K. Wong et al.}







\begin{abstract}
Biologically plausible Spiking Neural Networks (SNNs), characterized by spike sparsity, are growing tremendous attention over intellectual edge devices and critical bio-medical applications as compared to artificial neural networks (ANNs). However, there is a considerable risk from malicious attempts to extract white-box information (i.e., weights) from SNNs, as attackers could exploit well-trained SNNs for profit and white-box adversarial concerns. There is a dire need for intellectual property (IP) protective measures. In this paper, we present a novel secure software-hardware co-designed RRAM-based neuromorphic accelerator for protecting the IP of SNNs. Software-wise, we design a tailored genetic algorithm with classic XOR encryption to target the least number of weights that need encryption. From a hardware perspective, we develop a low-energy decryption module, meticulously designed to provide zero decryption latency. Extensive results from various datasets, including NMNIST, DVSGesture, EEGMMIDB, Braille Letter, and SHD, demonstrate that our proposed method effectively secures SNNs by encrypting a minimal fraction of stealthy weights, only 0.00005\% to 0.016\% weight bits. Additionally, it achieves a substantial reduction in energy consumption, ranging from $\times59$ to $\times6780$, and significantly lowers decryption latency, ranging from $\times175$ to $\times4250$. Moreover, our method requires as little as one sample per class in dataset for encryption and addresses hessian/gradient-based search insensitive problems. This strategy offers a highly efficient and flexible solution for securing SNNs in diverse applications\footnote{This code is available at: \url{https://github.com/u3556440/SNNGX_qSNN_encryption}.}.
\end{abstract}



\maketitle
\vspace{-6 pt}
\section{Introduction}
Spiking Neural Networks (SNNs) have emerged as a promising alternative to traditional Artificial Neural Networks (ANNs) in applications involving event data. Examples of biomedical applications include epilepsy detection~\cite{9401560}, brain-computer interfaces (BCIs)~\cite{kumar2022decoding}, and Alzheimer's disease detection~\cite{DOBORJEH2021522}. The advantages of SNNs, such as enhanced energy efficiency, inherent temporal dynamics, and biological plausibility~\cite{Yamazaki2022}, contribute to their potential for significantly advancing medical diagnostics and treatment in a wide range of contexts. 
As the application of SNNs continues to grow in critical domains, ensuring the intellectual property (IP) of SNN models becomes increasingly important. In this context, three unresolved SNN protection issues remain.


First, concerning model protection, SNN models are valuable assets due to the scarcity of neuromorphic training data and challenges associated with training. However, the protection of SNN model parameters has not been addressed. For instance, Abad \textit{et al.}~\cite{abad2023sneaky} demonstrated that traditional ANN defence methods, such as artificial brain stimulation and fine-pruning, were ineffective against Trojan-attacked SNNs but did not propose a defence solution for SNN weight protection. Moreover, Nagarajan~\textit{et al.} \cite{cryptography7020017} suggested decorrelating side-channel current with firing rate and constructing camouflage neurons could prevent side-channel attacks on SNN chips that read their weights; however, this cannot protect the weights against attackers via off-chip memory accesses~\cite{hua2018reverse}.


Second, SNN applications (e.g.,~\cite{9401560}) are subjected to various training schemes, including Spike Timing Dependent Plasticity (STDP)~\cite{FELDMAN2012556}, surrogate gradient~\cite{pmid29652587}, and ANN-to-SNN conversion training methods~\cite{rueckauer2017conversion}. Such diversity makes it difficult for traditional single-mode protection methods like gradient-based protection~\cite{8942041} to universally ensure model security. Moreover, the non-differentiability of leaky integrate-and-fire (LIF) neurons in SNNs poses challenges for gradient-based protection. For example, Cai~\textit{et al.}~\cite{8942041} found that ANNs could be encrypted up to about 20 bits per layer using Fast Gradient Sign Method (FGSM) to perturb top-k gradient weights against the encryption dataset. Nevertheless, when Liang~\textit{et al.}~\cite{pmid34473634} applied FGSM to SNNs for generating adversarial samples, they encountered gradient vanishing issues with zero gradient values outside the surrogate function window that makes gradient methods non-transferrable to SNNs.

\begin{figure*}[!thpb]%
    \centering
    {\includegraphics[width=1.9\columnwidth]{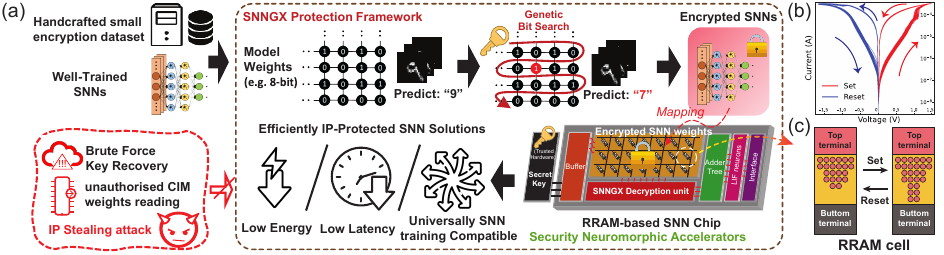}} \\
       \vspace{-10pt}  
    \caption{(a) SNNGX protection framework against IP stealing attack. (b) Quasi-static I-V sweeps of an RRAM cell demonstrate repeatable bipolar resistive switching behavior. (c) RRAM cell after set and reset operation.}%
    \label{fig:intro}%
       \vspace{-5pt}  
\end{figure*}

Third, in the ANN approach, white-box and black-box watermarking methods~\cite{uchida2017embedding,guo2018watermarking} are commonly employed to protect the IP of models. However, these methods necessitate a large number of sensitive training samples, which may contain private data, particularly in critical fields such as medical applications. Owing to privacy and legal concerns, these methods in real-world scenarios can face significant challenges.

To address the aforementioned challenges, we propose a universal protection framework \textit{SNNGX} -- using genetic algorithm-based XOR encryption for SNN protection and a novel XOR decryptor design on Resistive Random-Access Memory (RRAM) based accelerators to safeguard the IP of SNN parameters for massive reduction in hardware cost. The main innovations are as follows,
\vspace{-2pt}
\begin{itemize}

\item \textbf{The first IP protection scheme for SNNs.} We highlight the importance of IP protection for SNNs for the first time and propose a biologically inspired genetic algorithm-based XOR encryption solution that achieves SNN protection with extremely low hardware overhead.

\item \textbf{Universal and gradient-free SNN protection framework.} We introduce a black-box genetic algorithm for encryption which is not limited by specific SNN training algorithms and naturally circumvents SNN non-differentiability.

\item \textbf{No requirement for privacy-sensitive training data.} Extensive classification tasks on five event-driven or sequence datasets demonstrate that satisfactory protection can be achieved using only 1 encryption sample per category, significantly reducing data acquisition costs.

\item \textbf{ADC-less RRAM accelerator, zero additional latency.} We propose a novel architecture using RRAM for simultaneous decryption and in-memory computations, effectively eliminating any extra decryption latency. Moreover, this architecture lowers energy consumption by eliminating the need to write decrypted weights back to RRAM.

\end{itemize}
\vspace{-2pt}
The remaining paper is organized as follows: Sec.2 addresses the background and related work of SNN IP protection. Sec.3 discusses SNNGX protection principles. Sec.4 evaluates the simulation result of our study. Sec.5 is the conclusion.

\section{Preliminaries}
\subsection{SNN with LIF Neuron}
LIF neuron is a popular neuron model for various SNN applications \cite{9401560,kumar2022decoding, DOBORJEH2021522}, where the activation of each post-synaptic neuron $i$ in the network is governed by an RC circuit equation, \vspace{-4pt}
\begin{equation}
\ C_m \frac{\vartheta V_i(t)}{\vartheta t} = -g_{L}(V_i(t)-E_{L})+I_{ext} \vspace{-4pt}
\label{eq:lif}
\end{equation}
where $V_i(t)$ denotes membrane potential $V_{mem}$ of the post-synaptic neuron $i$ at time $t$, \(C_{m}\) refers to the capacitance constant of the membrane bi-lipid layer on which charges deposit. \(g_{L}\) and \(E_{L}\) represent membrane leakage conductance and leakage reversal potential separately. \(I_{ext}\) is the external pre-synaptic current. Bounded by a voltage threshold \(V_{th}\), the LIF neuron will fire a spike as an output $O_i(t)$ if membrane potential $V_i$ $\geq$ \(V_{th}\) at time $t$ that yields Eq.~(\ref{eq:spike}).
\begin{equation}
    O_i(t) =
    \begin{cases}
        1, & \text{if } V_i(t) \geq V_{th} \\
        0, & \text{if } V_i(t) < V_{th}
    \end{cases}
\label{eq:spike}
\end{equation}

\(I_{ext}\) equals the synaptic firing sum from all pre-synaptic neurons $j$. By simplifying the leaky term of LIF neurons to a single decay constant \(\lambda\), the LIF discrete dynamics can be approximated as,
\vspace{-4pt}
\begin{equation}
\ V_i(t+1) = \lambda \cdot V_i(t) (1-O_i(t)) +\sum_jW_{i,j} \cdot O_j(t)
\label{eq:d_LIF}
\vspace{-4pt}
\end{equation}
Unlike conventional ANNs, the non-differentiable nature of LIF neurons as in Eq.~(\ref{eq:spike}) divided SNNs into diverse training schemes. Unsupervised training of STDP ~\cite{FELDMAN2012556} adjusts weights by temporal difference between pre- and post-synaptic spikes in the forward pass. Supervised training of surrogate gradients ~\cite{pmid29652587} can be trained on rectangular, sigmoid or polynomial derivative approximations. ANN-to-SNN conversion avoids direct training on spiking neurons but training ReLU-activated ANN and migrates to IF neurons~\cite{rueckauer2017conversion}. This work tackles SNN diversity and performs universal protection. 
\vspace{-6 pt}

\subsection{IP Stealing and Protection}
Similar to ANNs~\cite{8942041}, well-trained SNNs are considered IP due to the investments in model design, training, and expert costs. However, SNN parameters implemented on the AI accelerator are at risk of being leaked. Early work from Clemens ~\textit{et al.}
\cite{helfmeier2013cloning} showed that using a Focused Ion Beam (FIB) workstation can mill away packaging and expose accelerators' underlying circuitry. The FIB workstation can cut and rewire area of interest to extract dynamic behaviour of an SRAM cell. Such risks are notably high for compute-in-memory (CIM) accelerators with emerging memory, as model information in storage devices is physically accessible even when the power is off~\cite{8942041, yan2019rram}. Microprobing with fully invasive decapsulation can provide an attacker access to any memory circuitry. 

Moreover, Nagarajan~\textit{et al.}~\cite{cryptography7020017} performs side-channel analysis on analog-hardware simulated SNN connecting with an external power source. They suggested SNN hyperparameters such as voltage threshold and neurons per layer, and synaptic weights can be estimated by correlation with average power over a fixed sampling window. Changes in these parameters affect spiking rate and time to spike value, which causes variation in power profile.

For IP protection in SNNs, Nagarajan~\textit{et al.}~\cite{cryptography7020017} defended power side-channel attack by varying the switching threshold of the neuron to decorrelate side-channel current with firing rate and constructing camouflage neurons. However, they did not address the white-box IP protection of weights with unauthorised off-chip memory accesses~\cite{hua2018reverse}. This motivates us to propose the first white-box IP protection method for SNNs. Abad~\textit{et al.}~\cite{abad2023sneaky} demonstrated that traditional ANN defence methods, like fine-pruning, were ineffective for trojan SNNs but did not propose a defence solution. Our work addresses this gap in limited SNN weight protection work.
\vspace{-6 pt}
\subsection{RRAM-based Neuromorphic Accelerator}

RRAM is a non-volatile CIM technology that stores information through resistance changes, which can emulate synaptic connections for neuromorphic computing~\cite{jo2010nanoscale,indiveri2013integration,lv2022variation}. As illustrated Fig. \ref{fig:intro}(b)(c), the set (red line) and reset (blue line) processes can program and erase the resistance state of the RRAM cell. During the set process, the redox reactions (such as silver or copper) and ion migration (depicted by the orange circles in the figure) leads to the formation of a conductive filament or filamentary structure. This establishes the memory cell's resistance state to a low-resistance level. Conversely, in the reset process, the dissolution of these conductive filaments or filamentary structures prompts the memory cell's resistance state to revert from a low-resistance level to a high-resistance level, consequently facilitating the erasure and rewriting of information. Through the set and reset processes, RRAM can achieve programmable non-volatile storage for the weights of SNN. RRAM-based neuromorphic accelerators have been an emerging solution due to high storage density and energy efficiency~\cite{wang2015energy}.

The non-volatile nature of RRAM allows attackers to easily access stored weights, posing a threat to the security of SNNs running on neuromorphic accelerators. In this work, we propose a novel, secure, ADC-less RRAM-based accelerator, as illustrated in Fig. \ref{fig:intro}. 


\subsection{Neural Network Bit Search}
A key issue in encrypting SNN is identifying the critical weights within the model, then conducting encryption. A possible solution is to determine weight sensitivity based on gradient used in ANNs \cite{8942041,rakin2019bitflip,hassibi1992second, frantar2022optimal,huang2024billm}. Disturbing elements that are more sensitive to gradients results in significant output deviations. The predicted functional change in the output error for encryption in original weight $\delta w$ is,
\vspace{-4pt}
\begin{equation}\label{gradient}
    \delta E = \mathop{\underline{(\frac{\vartheta E}{\vartheta w})^T \cdot
    \delta w}}\limits_{first-term} + \mathop{\underline{\frac{1}{2}\delta w^T \cdot\frac{\vartheta^2 E}{\vartheta w^2}\cdot\delta w}}\limits_{second-term} + O(||\delta w||^3)
\vspace{-4pt}
\end{equation}
where $w$ denotes the weights, 
the \textit{first-term} is the parameters' gradient, the \textit{second-term} is the second-order derivative. The third- and high-order terms are often ignored \cite{hassibi1992second}. Searching the largest indexes from the \textit{first-term} for encryption is a simple and fast strategy which is widely used \cite{8942041,rakin2019bitflip}. However, since pre-trained networks are at a local minimum error, the \textit{first-term} typically vanishes \cite{hassibi1992second}. Therefore, a more precise method than the normal gradient is to determine weight sensitivity based on the second-order derivative. This approach is widely used in network quantization and pruning \cite{hassibi1992second, frantar2022optimal,huang2024billm,huang2024slim}. In the \textit{second-term}, 
$\frac{\vartheta^2 E}{\vartheta^2 w}$ is equivalent to Hessian matrix, which can be calculated through strategic approximations \cite{hassibi1992second}.

However, due to the non-differentiable nature of LIF neurons in SNNs, the traditional first-order and second-order gradient determination methods widely applied in ANNs are constrained \cite{pmid34473634}. Therefore, we design a genetic bit search method and provide more detailed comparison in Sec. \ref{gradient_result}.


\section{SNNGX Protection Framework}
\subsection{Threat Model} 

Once the attacker obtains the weights of the SNNs, they could potentially clone it onto another neuromorphic chip. As a result, attackers may gain financial interests by reselling high-performance SNNs without authorization from the providers.

\textbf{Attacker's Capabilities.} Attackers can perform invasive circuitry modifications and microprobing on unauthorized RRAM-based accelerators. They can simply read the memory and extract the weight parameters, even without powering up the systems~\cite{chhabra2011nvmm}. They can also reverse-engineer the model through side-channel attacks \cite{cryptography7020017}. They may understand the principles of SNNGX.

\textbf{Attacker's Limitations.} Following literatures~\cite{guo2018puf,lin2020chaotic,linning24}, attackers cannot access private training samples and are unable to retrain unauthorized parameters. This is consistent with data protection laws stipulated in the EU GDPR \cite{8400247}, which require restrictive governance against unwanted surveillance and misuse of data targeting individuals. Furthermore, this is in line with the trade secrets of service providers that give their products a competitive advantage.


\subsection{Protection Overview} 
Our framework in Fig. \ref{fig:intro} addresses SNN IP security. SNNGX is divided into three parts: Genetic Bit Search, XOR Encryption, and XOR Decryption during inference. We first design a genetic bit search to destroy SNN accuracy with the minimum weight-bits XOR encrypted (bounded by reliability threshold in Sec. \ref{reliability}). The encrypted position described with key value `1' in a layer-wise sparse matrix becomes the "secret key" stored in a secure memory~\cite{seznec2010phase,saileshwar2018synergy,xin2017system}. This approach significantly reduces the storage capacity required for the key index compared to traditional encryption methods, marking a substantial improvement in both efficiency and resource utilization. An XOR decryption can be implemented on the CIM chip with respect to the "secret key" during inference. Our SNNGX decryption unit design is automatically deactivated when the key value is `0', thereby avoiding any extra power consumption and decryption latency. XOR encryption has been widely implemented in the security domain of DNN accelerators due to its hardware-friendly nature~\cite{huang2020xor,li2019p3m}.

\subsection{Genetic Bit Search \& XOR Encryption}
We employ a classic quantization method to convert full-precision weights into their $N_{bit}$ fixed-point representations \cite{rakin2019bitflip}.
To reduce the search space of vulnerable bits for XOR encryption, we only encrypt the most significant bit (MSB) of quantized weights, which empirically have a larger influence function on the loss than the remaining weights. A 1D sign bit weight vector $\boldsymbol{w_s^{raw}}\in \{-1,1\}^{n\cdot m}$ is created by flattening the extracted sign bit from the target layer of an $n \times m$  weight matrix,
\begin{equation}
\boldsymbol{w_{s}^{raw}} = (b_{sign}^{(0,0)},b_{sign}^{(0,1)},b_{sign}^{(0,2)} \ldots b_{sign}^{(n,m)}) 
\label{eq:raw_array}
\end{equation}

\textbf{\ding{182} Initial Population (see line 2 in Alg.1)}: A size of N adversarial $\boldsymbol{w_{s}'}$ populations are created for the evolutionary search. They are all flipped to maximum hamming distance with the original array to ensure low accuracy at the start and undergo mutation individually for better population diversity, 
\vspace{-2pt}
\begin{equation}
\forall \boldsymbol{w_{s}'} \in \boldsymbol{P_{init}}, \boldsymbol{w_{s}'} = mut(\boldsymbol{w_{s}^{raw}} \cdot -1)  
\vspace{-1pt}
\end{equation}
$\boldsymbol{P_{init}}$ refers to the initial population; function $mut(.)$ refers to the recovery mutation function, which will be covered in Eq.~(\ref{eq:mutaion}).

\begin{algorithm}
\caption{Genetic Bit Search \& XOR Encryption}
\setlength{\floatsep}{4pt}
\setlength{\textfloatsep}{4pt}
\begin{algorithmic}[1]
\Statex Input: 
\Statex \hspace{\algorithmicindent} $f_{\theta}$ (trained SNN), $\boldsymbol{D_{enc}}$ (tiny encryption dataset), 
\Statex \hspace{\algorithmicindent} $\epsilon = 50$ (NMNIST) (hamming distance bound), 
\Statex \hspace{\algorithmicindent} $G=120$ (\# of Gen), $p_m=0.05$ (mutation probability), 
\Statex \hspace{\algorithmicindent} $N=100$ (Population Size), $r=60\%$ (Elites retaining rate)
\Statex Output:
\Statex \hspace{\algorithmicindent} $f_{E(\theta)}$ (Encrypted SNN model)
\State $\textbf{function}$ main() 
\State $Population\leftarrow Init()$ //Max. Hamming distance
\For {$gen=1,2,\ldots,G$}
    \State $new Pop\leftarrow\emptyset$ 
    \State evaluate $Population$ //Accuracy \& Hamming distance 
    \State select best $r\%$ $Population$ to $newPop$
    \State split $Population$ to males and females
    \While {places left in $new Pop$}
        \State $child1,child2\leftarrow$ Crossover($parent1,parent2$)
        \State //recover some bits for $newPop$
        \State $\boldsymbol{w_{s,1}'}\leftarrow Mutation_{Rcry}(\boldsymbol{w_{s,1}'} \in newPop)$
        \State $\boldsymbol{w_{s,2}'}\leftarrow Mutation_{Rcry}(\boldsymbol{w_{s,2}'} \in newPop)$
        \State add $child1,child2$ to $new Pop$
        \EndWhile
        \State $Population\leftarrow newPop$ 
\EndFor
\State $f_{E(\theta)}\leftarrow$ update $f_{\theta}$ with $Secret$ $Key$ $(\boldsymbol{w_{s}^{best}} \oplus \boldsymbol{w_{s}^{raw}})$
\State \textbf{return} $f_{E(\theta)}$, $d(\boldsymbol{w_{s}^{best}},\boldsymbol{w_{s}^{raw}})$, $\lvert \boldsymbol{w_s} \rvert$, $Secret$ $Key$ 
\end{algorithmic} 
\end{algorithm}

\textbf{\ding{183} Fitness Function (see line 5 in Alg.1)}:
The minimization of fitness function is solved by the following constrained optimization problem,
\vspace{-4pt}
\begin{equation}
\operatorname*{min}_{\boldsymbol{w_{s}'}} L(\boldsymbol{w_{s}'})  \hspace{1em}
 subject \hspace{0.5em} to \hspace{1em} D(\boldsymbol{w_{s}'}) \leq \epsilon
 \label{eq:fitness}
 \vspace{-4pt}
\end{equation}
where in Eq.~(\ref{eq:fitness}), the loss function $L(\boldsymbol{w_{s}'})$ returns a loss value $l$ by updating the weight from XOR-processing the sign bit of the $N_{bit}$ original layer weight with the encryption position key $\boldsymbol{w_{s}'} \oplus \boldsymbol{w_{s}^{raw}}$ reshaped to $n \times m$, a.k.a the \textbf{XOR encryption process}, and forward pass with a small testing data subset $\boldsymbol{D_{enc}} \in \{(x,y)\}^{z}$ such that $z \ll$ number of testing samples to return an accuracy. The distance function $D(\boldsymbol{w_{s}'})$ returns the hamming distance $d$ between the current input array $\boldsymbol{w_{s}'}$ and the original array $\boldsymbol{w_{s}^{raw}}$ bounded with a distance budget $\epsilon$, so Eq.~(\ref{eq:fitness}) can be rewritten as,
\vspace{-0pt}
\begin{equation}
\operatorname*{min}_{\boldsymbol{w_{s}'}} fitness(L(\boldsymbol{w_{s}'}),D(\boldsymbol{w_{s}'})) 
\vspace{-4pt}
\label{eq:minF}
\end{equation}

The fitness score takes in hamming distance $d$ and loss value $l$ from encryption data $\boldsymbol{D_{enc}}$. Since a good "encrypted" model should have both low precision and minimal bits affected, the score has to be minimized to evaluate the best $\boldsymbol{w_{s}'}$ from the current generation with the score calculated by the following equation,
\begin{equation}
    fitness(l,d) = 
    \begin{cases}
        \epsilon \cdot l, & \text{if } d \leq \epsilon \\
        d + d \cdot l, & \text{if } d > \epsilon 
    \end{cases}
\end{equation}

The fitness score is first forced to reduce the hamming distance $d$ when $d$ is larger than the distance budget $\epsilon$. The score stops measuring the distance component when it is within the desired boundary and prioritises vulnerable bit combinations.

\textbf{\ding{184} Selection (see line 6 in Alg.1)}: All $\boldsymbol{w_{s}'} \in \boldsymbol{P}$ are re-ranked in an ascending order with respect to their fitness score. Only $r$\% stimuli with the lowest score are preserved in the next generation while the rest stimuli are removed.


\textbf{\ding{185} Crossover (see line 9 in Alg.1)}:
Two random $\boldsymbol{w_{s}'} \in \boldsymbol{P} $ are chosen to be parents and inter-swapped with a uniform crossover that produces two children sign bit vectors. The uniform crossover operation is simple yet provides better recombination potential \cite{pmid33162782} such that each bit in the $\boldsymbol{w_{s}'}$ is treated independently instead of inter-swapping a large segment of bits like K-point crossover.

\textbf{\ding{186} Recovery Mutation (see lines 11-12 in Alg.1)}: Some bits from $\boldsymbol{w_{s}'}$ are partially recovered to the original sign bit vector $\boldsymbol{w_{s}^{raw}}$ by recovery mutation function $mut(.)$. A flipping filter $\boldsymbol{p}\in\{0,1\}^{n\cdot m}$ is created with a binomial distribution $X \sim B((n\cdot m),p_m)$. A conditional function $\delta(\boldsymbol{x}\neq \boldsymbol{w_{s}^{raw}})$ returns 1 if two sign bit vectors are unequal element-wise and 0 otherwise,
\vspace{-0pt}
\begin{equation}
mut(\boldsymbol{x}) = \boldsymbol{x} \odot (2(\delta(\boldsymbol{x}\neq \boldsymbol{w_{s}^{raw}}) \odot \boldsymbol{p}) -1) \cdot -1
\label{eq:mutaion}
\vspace{-2pt}
\end{equation}

The expected bits recovered during each generation are approximately $n\cdot m \cdot p_m$. Therefore, we can approximate the number of generations $G$ needed for converging $\boldsymbol{w_{s}'}$ to the distance budget $\epsilon$ with $\boldsymbol{w_{s}^{raw}}$ by Eq.~(\ref{eq:generation}). We empirically introduce a reservoir factor of 135\% to ensure the complete convergence of fitness score. 
\begin{equation}
G \approx \lceil \frac{ln(\epsilon)-ln(n\cdot m)}{ln(1-p_{mut})} \cdot 135\% \rceil
\label{eq:generation}
\end{equation}

\textbf{\ding{187} Final Population (see line 17 in Alg.1)}:
After $G$ generations, the individual $\boldsymbol{w_{s}'}$ with the lowest fitness score from the population pool is picked as $\boldsymbol{w_{s}^{best}}$. 
A position array, a.k.a the final "secret key" to decrypt the encrypted model, is returned from the genetic algorithm by $\boldsymbol{w_{s}^{best}} \oplus \boldsymbol{w_{s}^{raw}} $. The sign bit of the target layer in a trained $N_{bit}$ SNN model $f_{\theta}$ is therefore XOR-processed with the reshaped secret key to become an \textbf{XOR encrypted model} $f_{E(\theta)}$.

\begin{figure}[!h]%
    \centering
    \vspace{-8pt} 
    {\includegraphics[width=0.95\columnwidth]{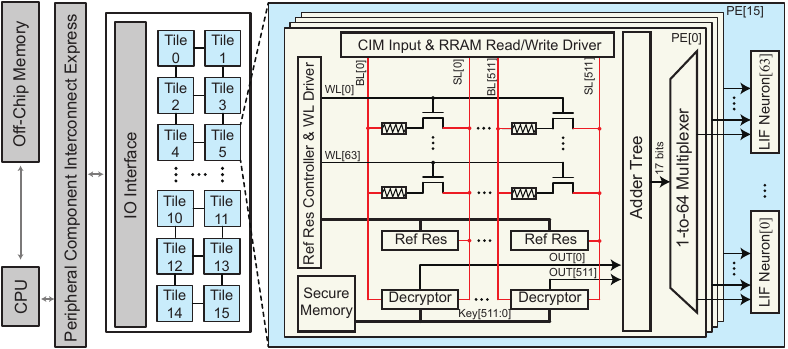}} \\
       \vspace{-10pt}  
    \caption{The schematic of SNNGX accelerator architecture.}%
    \label{fig:schematic2}%
       \vspace{-10pt}  
\end{figure}

\subsection{RRAM-based Accelerator Architecture}

To minimize latency and energy consumption during the decryption in the inference phase, we propose a novel architecture using RRAM for simultaneous decryption and in-memory computations, as illustrated in Fig.~\ref{fig:schematic2}, which is tailored to eliminate any additional decryption latency during inference. 

The architecture illustrates a traditional CPU computing module on the left flank, which, upon initiation of SNN inference, interfaces with a purpose-engineered chiplet via the Peripheral Component Interconnect Express (PCIe) protocol. This chiplet is architecturally segmented into 16 tiles. Each tile is an ensemble of 64 digital LIF neurons and 16 Processing Elements (PEs) with 32$\times$1024 RRAM cells in a PE. An in-depth exposition of the structural and computational principles of a singular PE is presented forthwith.

In the PE, Word lines (WLs) are governed by a WL Driver, which is constituted by a 6-to-64 decoder and transmission gates. During reading, writing, and computing, the isolation of RRAM between rows is ensured by the selective activation of a WL, thereby mitigating inter-row interference. Correspondingly, bit lines (BLs) and source lines (SLs) are managed by BL and SL Drivers, respectively. 
The Decryptor serves a pivotal role in key input and the modulation of the precharge (PRE) signal, which is intrinsically sensitive to the key signal extracted from secret memory. Should the ascending edge of a cycle discern a key value of `1', the PRE is reset to `0'; conversely, in the absence of this condition, the PRE signal is maintained at `1'.

\begin{figure}[!t]%
    \centering
    {\includegraphics[width=1\columnwidth]{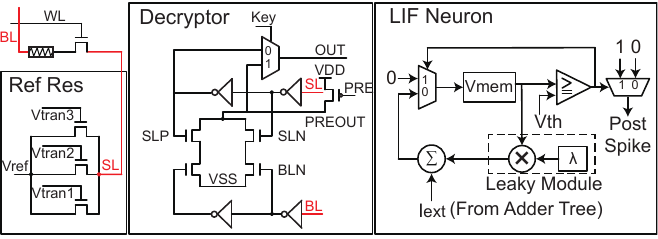
    }} \\
       \vspace{-10pt}  
    \caption{The schematic of reference resistor (left panel), decryptor (middle panel) and LIF Neuron (right panel).}%
    \label{fig:schematic3}%
       \vspace{-16pt}  
\end{figure}

In Fig.~\ref{fig:schematic2}, the reference resistor block (Ref Res) is detailed in Fig.~\ref{fig:schematic3} (left panel). The external equivalent resistance of the Ref Res is fine-tuned through the modulation of voltages across $V_{tran3}$, $V_{tran2}$, and $V_{tran1}$. This dynamic adjustment of the reference resistor is used to identify the optimal reference resistor. Such calibration is necessary to align with the dynamic resistance range of RRAM.

A weight value is quantized into an 8-bit binary programmed into eight RRAM cells, where a reset (set) operation transitions the RRAM into an HRS (LRS) for encoding a `0' (`1').

In the computational process of a singular bit element within our proposed architecture, a departure is made from the conventional approach of decrypting the encrypted weight prior to multiplication. Instead, leveraging logical equivalence, we follow "Multiplication first, Decryption afterward", as delineated in Eq.~(\ref{eq:distributive}),
\vspace{-5pt}

\begin{equation}
x \cdot (w^E \oplus k) :\iff (x \cdot w^E) \oplus x 
\label{eq:distributive}
\vspace{-3pt}
\end{equation}
\vspace{-3pt}
\renewcommand\qedsymbol{ }
\vspace{-3pt}
\begin{proof}
\vspace{-3pt}
\begin{equation*}
(x \cdot w^E) \oplus x
\vspace{-3pt}
\end{equation*}
\begin{equation*}
\iff ((x \land w^E) \land \lnot x) \lor (\lnot(x \land w^E) \land x)
\vspace{-3pt}
\end{equation*}
\begin{equation*}
\iff (\lnot x \lor \lnot w^E) \land x
\vspace{-3pt}
\end{equation*}
\begin{equation*}
\iff x \land \lnot w^E
\vspace{-3pt}
\end{equation*}
\begin{equation*}
\iff x \cdot (w^E \oplus k) 
\vspace{-3pt}
\end{equation*}
\end{proof}
\vspace{-15pt}
Here, $x$ denotes the single-bit input, $w^E$ represents the single-bit encrypted weight and $k$ stands for the Key = `1'. The symbols $\lnot$, $\land$, and $\lor$ correspond to the logical `not', `and', and `or'.

\begin{table}[!h]
    \centering
    \vspace{-5pt}
    \caption{Truth table for $w^E$ multiplication calculations}
    \vspace{-10pt}
    \label{tab:calculate}
    \scalebox{1}{%
    \setlength{\tabcolsep}{1.8pt} 
    \begin{tabular}{cccc|cccc|c}
    \hline
    Weight & Key &$w^E$&  $x$ (BL) & SL    & SLP  & SLN  & BLN     & OUT\\
    \hline\hline
     \cellcolor[HTML]{EFEFEF}1   & \cellcolor[HTML]{EFEFEF}1   &\cellcolor[HTML]{EFEFEF}0(HRS) &  \cellcolor[HTML]{EFEFEF}1(1.2V)   & \cellcolor[HTML]{EFEFEF}0.66V & \cellcolor[HTML]{EFEFEF}0V   & \cellcolor[HTML]{EFEFEF}1.8V &\cellcolor[HTML]{EFEFEF}0V   & \cellcolor[HTML]{EFEFEF}1(1.8V)\\
      1      & 1   &0(HRS) &  0(0.6V)   & 0.6V  & 0V   & 1.8V & 1.8V & 0(0V)\\
      \cellcolor[HTML]{EFEFEF}0      & \cellcolor[HTML]{EFEFEF}1   &\cellcolor[HTML]{EFEFEF}1(LRS) &  \cellcolor[HTML]{EFEFEF}1(1.2V)   & \cellcolor[HTML]{EFEFEF}1.14V &\cellcolor[HTML]{EFEFEF} 1.8V & \cellcolor[HTML]{EFEFEF}0V   & \cellcolor[HTML]{EFEFEF}0V   &\cellcolor[HTML]{EFEFEF}0(0V)\\
      0      & 1   &1(LRS) &  0(0.6V)   & 0.6V  & 0V   & 1.8V & 1.8V & 0(0V)\\

    \hline
    \end{tabular}
    }
    
    \vspace{-7pt}  
\end{table}

Table~\ref{tab:calculate} presents the truth table for the multiplication of a single-bit encrypted weight ($w^E$). `Weight' denotes the original weight and `Key' signifies the encryption key. The subsequent terms, BL, SL, SLP, SLN, BLN, and OUT refer to the voltages observed during the decryption process which are explicitly labeled in the Decryptor as shown in Fig.\ref{fig:schematic3} (middle panel). Taking the first row as an example, where the true weight is `1' and undergoes an XOR operation with a key value of `1', the $w^E$ becomes `0', resulting in the RRAM adopting an HRS with a resistance of 100K$\Omega$. When an input of `1' is applied, a voltage of 1.2V is introduced through the BL. Simultaneously, the external equivalent resistance of the Ref Res is maintained at 10K$\Omega$, generating a reference voltage (V$_{\text{ref}}$) of 0.6V. As the WL voltage attains 1.8V, the NMOS switch is triggered, leading to an SL voltage of 1.14V owing to the differential voltage generated by the partial pressures in the RRAM and Ref Res. For the Decryptor in Fig.~\ref{fig:schematic3} (middle panel), the threshold voltage of the custom-designed inverter is set to 0.9V, with a margin of about 0.1V. Consequently, after passing through two inverters, the SLP outputs the final calculated result of `1' (1.8V).

In the decryptor, the activation of the PRE signal depends on the assertion of Key. If the Key is `0', indicating that the current weight is not encrypted, the output is the SLP signal. During decryption (when key = 1), the PREOUT line is precharged to 1.8V while the XOR operation is done by four NMOS transistors. If both NMOS transistors on the same side are active concurrently, the PREOUT line will be pulled down to 0V. If not, it will remain at 1.8V.

The outputs from each decryptor are concurrently fed into the Adder Tree, capable of aggregating eight 8-bit signed binary numbers within a single cycle. Subsequently, the aggregated results from multiple PEs are directed to the corresponding digital LIF neuron through their individual 1-to-64 multiplexers. 

\vspace{-2pt} 
\subsection{Digital LIF Neuron }
Our designed digital LIF neuron is depicted in Fig.~\ref{fig:schematic3}, which expresses discrete LIF dynamics in Eq. ~(\ref{eq:d_LIF}). The circuit comprises four components and two multiplexers (MUXs). The first one is on-chip SRAMs for storing membrane potential $V_{mem}$. The second is a comparator to compare $V_{mem}$ with predefined $V_{th}$. Once $V_{mem} \geq V_{th}$, the comparator controls the post-spike MUX and reset MUX respectively to generate a post-spike and reset the membrane potential. The third is leaky module which performs a fixed-point multiplication of the $V_{mem}$ with leaky factor $\lambda$ to achieve membrane potential leakage function. The fourth is the adder that updates $V_{mem}$ for the subsequent time step by adding weighted input sum $I_{ext}$ from adder tree and leaky $V_{mem}$ at the present time step.
\vspace{-2pt} 
\subsection{Brute Force Key Recovery}
Encrypting a minimum number of bits in a model not only brings efficiency improvement but also may raise brute-force breakable problems. We answer the minimum number of weights needed to ensure reliability for the most efficient (worst-case) scenario. This follows Kerckhoffs's principle~\cite{6769090}, which states that attackers may understand SNNGX encryption method except for secret keys.

\vspace{+2pt} 
\textbf{Most efficient decryption case.}
We assume the deployment of SNNGX be on the least-weight layer of SNN and also a very small number of weights encrypted (e.g., single digit). The time complexity for all recoverable sign-bit in one layer with $n$ weights is $O(2^n)$, which is usually impossible to traverse.

The intelligent attacker may start trial and error from the least weight layer and recover the sign bit from flipping 1 to k bits, where k is limited to the attacker's computing power. Such a bit-flip recovery attack can be interpreted as the sum of the binomial coefficient on the least weight layer from 1 to k with $n$ weights,
\vspace{-6pt}
\begin{equation}
\frac{n!}{1!(n-1)!} + \frac{n!}{2!(n-2)!} +... + \frac{n!}{k!(n-k)!} =\sum^{k}_{i=1}\binom{n}{i}
\vspace{-4pt}
\label{eq:binomial}
\end{equation}

For the upper bound in Eq.~(\ref{eq:binomial}), we could derive the time complexity bounded with geometric series approximation as follows,
\vspace{-6pt}
\begin{equation*}
\frac{\sum^{k}_{i=0}\binom{n}{i}}{\binom{n}{k}} =\frac{\binom{n}{0}}{\binom{n}{k}} +\frac{\binom{n}{1}}{\binom{n}{k}} + \frac{\binom{n}{2}}{\binom{n}{k}} + ... +\frac{\binom{n}{k-1}}{{\binom{n}{k}}}+ \frac{\binom{n}{k}}{{\binom{n}{k}}}
\end{equation*}
\begin{equation*}
= \frac{k!(n-k)!}{n!} +\frac{k!(n-k)!}{(n-1)!} + \frac{k!(n-k)!}{2!(n-2)!}  + ... +\frac{k!(n-k)!}{(k-1)!(n-k+1)!}  +1
\end{equation*}
\begin{equation}
= 1+ \frac{k}{n-k+1}  + \frac{k(k-1)}{(n-k+1)(n-k+2)} + ... + \frac{k!}{\prod^{k}_{i=1}(n-k+i)}
\label{eq:gs}
\end{equation}

For $k < (n+1)/2$,  Eq.~(\ref{eq:gs}) can be bounded with geometric series, which is also a tight bound considering attack feasibility for $k << n$,
\vspace{-4pt}
\begin{equation*}
\frac{\sum^{k}_{i=0}\binom{n}{i}}{\binom{n}{k}} \leq 1+ \frac{k}{n-k+1}  + (\frac{k}{n-k+1})^{2} + ... + (\frac{k}{n-k+1})^{k}
\vspace{-4pt}
\end{equation*}
\begin{equation}
\vspace{-4pt}
\leq \frac{n- (k-1)}{n-(2k-1)}
\vspace{-4pt}
\end{equation}
\begin{equation}
\sum^{k}_{i=1}\binom{n}{i} \leq \binom{n}{k}\frac{n-k+1}{n-2k+1} -1
\vspace{-2pt}
\end{equation}

As a result, the time complexity for an intelligent attacker can be more realistically estimated with $O(\binom{n}{k}\frac{n-k+1}{n-2k+1})$ instead of $O(2^n)$. For example, we encrypt the input layer of LeNet5 consisting of 1 in-channel and 6 out-channels with $5 \times 5$ kernels of total $n=150$ weights. The brute force key search complexity of traversing all combinations $O(2^n)$ is $1.42\times 10^{45}$. However, brute force key searching up to 5 among 150 weights reduces the search time to $O(\binom{n}{k}\frac{n-k+1}{n-2k+1})$ which is $6.13\times 10^{8}$. It is feasible for full key recovery if only 5 bits are encrypted for LeNet5 against $O(\binom{n}{k}\frac{n-k+1}{n-2k+1})$ attack.

\vspace{+2pt}
\textbf{Danger of partly recovered key.} Even though the intelligent attacker may not recover the full key from a brute force attack, the attack may still threaten the model IP with a partly recovered key on important weights, which is studied in Sec. 4.4. 


\section{Simulation Results}
\subsection{Experimental Setup}
\textbf{Benchmarks and baselines.}
Experiments are evaluated on a digital computer with an AMD Ryzen 9 7950x CPU and an NVIDIA GeForce RTX 4090 GPU.
This paper deploys the SNNGX framework on 5 different datasets, and the hyperparameters of SNNGX used in the experiments of this paper are shown in Table~\ref{tab:hyper}.
Datasets modalities across event images (NMNIST\cite{fnins.2015.00437} \& DVSGesture\cite{8100264}), biomedical sequence signals (EEG Motor Movement/Imagery Dataset (EEGMMIDB)) \cite{pmid15188875}, tactile data (Braille letter) \cite{mullercleve2022braille}, and event speech (Spiking Heidelberg Dataset (SHD)) \cite{9311226} to verify its scalability on various SNN architectures (see Table~\ref{tab:Modelresult} for details). All SNN model weights are quantized to 8-bit, following the method mentioned in~\cite{rakin2019bitflip}. 
The area, latency, and power consumption of the analog portion of the entire PE are evaluated using the complete schematic and layout in Cadence Virtuoso; digital LIF neuron part is assessed using the Synopsis Design Compiler's synthesis tools. 

\begin{table}[!h]
\centering
\vspace{-6pt}
\caption{Hyper-parameter settings for SNNGX.}
 \vspace{-10pt}
 \resizebox{1\linewidth}{!}{
\begin{tabular}{lccccc}
\toprule
& NMNIST & DVSGesture & EEGMMIDB &  Braille Letter & SHD\\
\toprule\toprule
\cellcolor[HTML]{EFEFEF} $\epsilon$ (Distance bound) & \cellcolor[HTML]{EFEFEF} 50 & \cellcolor[HTML]{EFEFEF}20 & \cellcolor[HTML]{EFEFEF}80 & \cellcolor[HTML]{EFEFEF}20 & \cellcolor[HTML]{EFEFEF}250 \\

\midrule

$G$ (\# of Generation)  & 120 & 108 & 55 & 190 & 72\\

\midrule
\cellcolor[HTML]{EFEFEF}$N$ (\# of Population) &\cellcolor[HTML]{EFEFEF} 100 & \cellcolor[HTML]{EFEFEF}100 &\cellcolor[HTML]{EFEFEF}100 &\cellcolor[HTML]{EFEFEF}100 &\cellcolor[HTML]{EFEFEF}100  \\

\midrule
$p_m$ (mutation rate) & 0.05 & 0.05 & 0.05 & 0.05 & 0.05  \\

\midrule
\cellcolor[HTML]{EFEFEF}$r$ (retain rate) &\cellcolor[HTML]{EFEFEF} 0.6 & \cellcolor[HTML]{EFEFEF} 0.6 &\cellcolor[HTML]{EFEFEF}0.6 &\cellcolor[HTML]{EFEFEF}0.6 & \cellcolor[HTML]{EFEFEF}0.6 \\

\bottomrule
\end{tabular}}
\label{tab:hyper}
 \vspace{-10pt}
\end{table}
\textbf{Evaluation Metrics.}
For IP protection validation, similar to the literatures in that of ANNs~\cite{8942041,lin2020chaotic,guo2018puf}, we use validation accuracy after encryption as a benchmark. Lower (higher) accuracy implies better (poor) IP protection. Regarding hardware efficiency, lower (higher) latency and smaller (larger) energy consumption indicate better (worse) efficiency.

\begin{figure}[!t]%
    \centering
    {\includegraphics[width=1.\columnwidth]{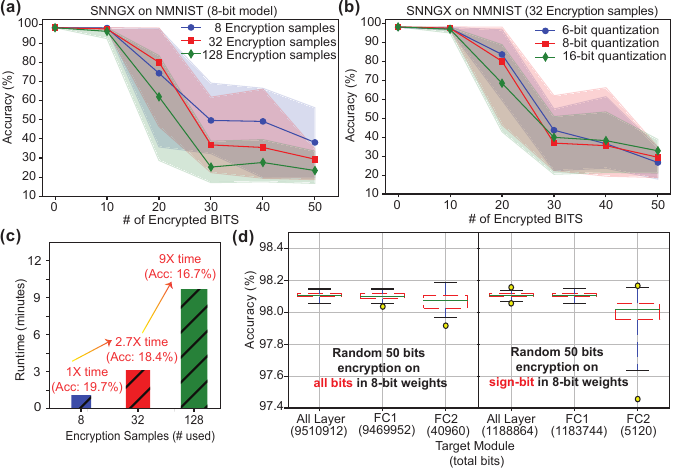}} \\
       \vspace{-10pt}  
    \caption{Accuracy of NMNIST (output layer with genetic encryption) by varying number of encryption samples (a) and by varying bits for quantization (b). Genetic encryption time cost (c). Baseline of random encryption with 50 bits 100 times using 8-bit encryption vs. sign-bit encryption (d).}%
    \label{fig:nmist}%
    \vspace{-16pt}  
\end{figure}
 \vspace{-3pt}
\subsection{IP Protection}
\subsubsection{Encryption samples and quantization bitwidth.}
We first study the impact of encryption dataset size (i.e., 8/32/128) and weight quantization bitwidth (i.e., 6/8/16 bits) on the performance of SNNGX on NMNIST. Fig.~\ref{fig:nmist}(a) and (b) use 10\% minimum and maximum outliers filtered result for visualization purposes. Both Fig.~\ref{fig:nmist}(a) and (b) show that using only 30 to 50 bits with SNNGX, we can destroy the accuracy of the well-trained model on NMNIST and attain an anti-IP stealing effect. Fig.~\ref{fig:nmist}(a) suggests that using more encryption samples improves encryption stability. Interestingly, however, considering 8 samples are smaller than the total number of 10 classes of NMNIST, it has a decent accuracy lower bound which is close to the case of 128 encryption samples using 40 and 50 encrypted bits. This suggests our SNNGX encryption is data-efficient, where oversampling of encryption dataset shall not affect the protection performance considerably. Whereas in Fig.~\ref{fig:nmist}(b), varying the number of bit quantization likely does not affect the performance of our framework because we only encrypt the sign bits of SNN weights. Fig.~\ref{fig:nmist}(c) demonstrates the encryption time cost consumed in Fig.~\ref{fig:nmist}(a). Time grows exponentially with the uniform 4X increase of samples. In fact, the encryption time of SNNGX is dominated by forward pass computational complexity for total operations within the network, number of populations $N$ and search space for fitness function to converge can affect though. This time cost estimation can be extended from NMNIST to other models. For the ablation study, we measure the model performance with random bit encryption. Fig.~\ref{fig:nmist}(d) shows the impact of random 50-bit encryption on the NMNIST dataset in 2 scenarios: 8-bit random encryption and sign-bit random encryption. The classification accuracy is slightly decreased, which corroborates the genetic bit search plays a critical role in identifying bits capable of protecting SNN IPs.

\vspace{-2pt}
\subsubsection{Vulnerable layer to encrypt.} 
\begin{figure}[!t]%
    \centering
    {\includegraphics[width=1\columnwidth]{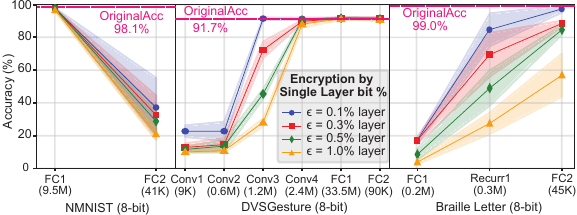}} \\
       \vspace{-10pt}  
    \caption{NMNIST: SNN(Left), DVSGesture: CSNN (Mid) and Braille Letter: RSNN (Right) by different single layer encryption performance}%
    \label{fig:by_layer}%
       \vspace{-12pt}  
\end{figure}

In Fig.~\ref{fig:by_layer}, we encrypt every layer 5 times for SNN, CSNN and RSNN separately to show the best layer for efficient encryption in different architectures. We try 4 efficiency levels, 0.1\%, 0.3\%, 0.5\% and 1\% of 8-bit layer bits. The $\epsilon$ is decided by number of picked layer weights $\times$ quantization bit-width $\times$ efficiency level. We observe that the first few input layers and the output layer are the most vulnerable. 1st and 2nd layers are usually more vulnerable for efficient encryption unless number of input layer weights $>>$ output layer. 0.1\% to 1\% of the picked layer bits is a good estimation for efficiency level.

\definecolor{lightgray}{gray}{0.9}

\begin{table*}[!thbp]
\centering
\caption{Protection results across various datasets and SNN architectures.}
 \vspace{-10pt}
 \resizebox{0.9\linewidth}{!}{
\begin{tabular}{lccccc}
\toprule
& NMNIST & DVSGesture & EEGMMIDB&  Braille Letter & SHD \\
\toprule\toprule
8-bit SNN Accuracy (\%) & 98.1 & 91.7 & 86.2 & 99.0 & 65.6 \\
\midrule
SNN Architecture\textsuperscript{*} & 2312F-512F-10F & 4CB-4096F-1024F-11F & 3CB-3TC-256R-2F & 128F-200R-28F & 700F-200R-20F \\


\midrule
Model Total Bits & 9,510,912 Bits & 37,782,528 Bits & 5,579,264 Bits & 569,600 Bits & 1,472,000 Bits \\

\midrule
Number of Classes & 10 & 11 & 2 & 28 & 20 \\

\midrule
Encryption-samples for SNNGX & 8 & 16   & 4  & 32  & 32  \\
\midrule
\cellcolor{lightgray} SNNGX: Encrypted Bits & \cellcolor{lightgray} 37 Bits & \cellcolor{lightgray} 17 Bits & \cellcolor{lightgray} 77 Bits & \cellcolor{lightgray} 10 Bits & \cellcolor{lightgray} 230 Bits \\
\cellcolor{lightgray} SNNGX:  Accuracy (\%) & \cellcolor{lightgray} 19.7 & \cellcolor{lightgray} 9.1  &  \cellcolor{lightgray} 48.9 &\cellcolor{lightgray} 26.4 & \cellcolor{lightgray}19.6 \\
\cellcolor{lightgray} Encryption Ratio (Model-wise) (\%) & \cellcolor{lightgray} \textbf{0.00039} & \cellcolor{lightgray} \textbf{0.000045}  & \cellcolor{lightgray} \textbf{0.0014} & \cellcolor{lightgray} \textbf{0.0018} & \cellcolor{lightgray} \textbf{0.016} \\
\midrule

\cellcolor{lightgray} Random-bit Encryption: Encrypted Bits & \cellcolor{lightgray} $8.2\times10^{5}$ Bits & \cellcolor{lightgray} $1.7\times10^{6}$ Bits & \cellcolor{lightgray} $1.7\times10^{6}$ Bits & \cellcolor{lightgray} $7.0\times10^{4}$ Bits & \cellcolor{lightgray} $2.0\times10^{5}$ Bits \\

\cellcolor{lightgray} Random-bit Encryption: Accuracy (\%) & \cellcolor{lightgray} 19.5 & \cellcolor{lightgray} 9.1 & \cellcolor{lightgray} 49.6 & \cellcolor{lightgray} 25.1  & \cellcolor{lightgray} 18.8\\

\cellcolor{lightgray} Encryption Ratio (Model-wise) (\%) & \cellcolor{lightgray} 8.6 & \cellcolor{lightgray} 4.5 & \cellcolor{lightgray} 30.5 & \cellcolor{lightgray} 12.3 & \cellcolor{lightgray} 13.6 \\

\bottomrule
\end{tabular}
}
\begin{tabular}{@{}p{\textwidth}@{}}
\hspace{8em}\footnotesize{\textsuperscript{*}F is Fully Connected; CB stands for Convolutional Block; TC represents Temporal Convolutional Block; R is Recurrent Layer.}
\end{tabular}
\label{tab:Modelresult}
 \vspace{-18pt}
\end{table*}

\vspace{-2pt}
\subsubsection{Categorical accuracy breakdown.} 
\begin{figure}[!t]%
    \centering
    {\includegraphics[width=1\columnwidth]{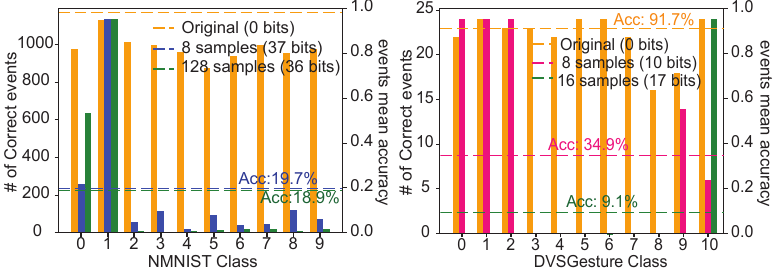}} \\
       \vspace{-10pt}  
    \caption{Categorical accuracy and undersampling of encryption dataset for NMNIST (Left) and DVSGesture (Right).}%
    \label{fig:attack}%
       \vspace{-16pt}  
\end{figure}

The class-wise classification accuracy of the protected model is shown in Fig.~\ref{fig:attack}, for both NMNIST (8/128 encryption samples) and DVSGesture (8/16 encryption samples) datasets. For both datasets, we observe that the class-wise accuracy of most classes experiences a significant drop. The model tends to classify encryption samples into a few classes, such as class 1 in the NMNIST and class 10 in the DVS gesture. Meanwhile, Fig.~\ref{fig:attack} also investigates the undersampling of NMNIST and DVSGesture encryption datasets which have 10 and 11 classes respectively. We suggest getting at least one encryption datapoint per class for best encryption result, which is 10 and 11 samples for NMNIST and DVSGesture. 2 and 3 undersampled encryption leave them approximately 2 and 3 classes operating normally.

\vspace{-2pt}
\subsubsection{IP protection generalization.} Table~\ref{tab:Modelresult} shows the result for IP protection over all 8-bit SNN models of different modalities and architectures. By using 10 bits to 230 bits for encryption, all models' accuracy drops to near-random accuracy. This table only aims to show how extremely SNNGX is able to encrypt models at the fewest bits. However, robustness against brute force recovery attacks is yet to be proven, which will be covered in Sec. 4.4. The choice of layer for encryption in Table~\ref{tab:Modelresult} is justified in Sec. 4.2.2, where input layer is selected unless output layer has significantly fewer parameters. We also compare our SNNGX with random bit encryption, which shows that SNNGX is 870$\times$--100,000$\times$ more efficient than model-wise random bit encryption in terms of bits used to achieve the same level of protection.


\begin{figure}[!t]%
    \centering
    {\includegraphics[width=1\columnwidth]{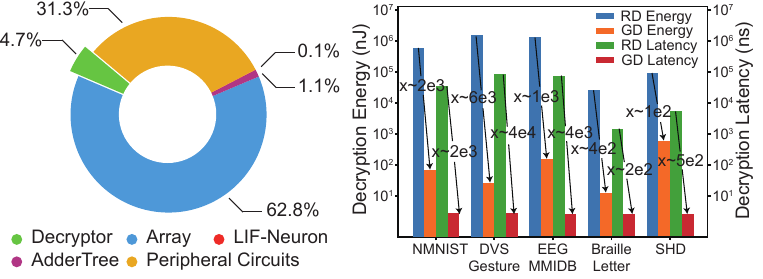}} \\
       \vspace{-10pt}  
    \caption{The area breakdown and the comparison of energy consumption and decryption latency between RD and GD.}%
    \label{fig:hardwarecost}%
       \vspace{-20pt}  
\end{figure}

\vspace{-4pt}
\subsection{Hardware Cost}
In Fig.~\ref{fig:hardwarecost}, the left side shows the area breakdown of the entire architecture, with the decryption module accounting for just $4.7\%$ of the total area. On the right side, we compare the energy consumption and decryption latency of traditional encryption and decryption methods in conventional RRAM-based architecture with our approach. The conventional random decryption (RD) method employs random-bit encryption for network security and rewrites the RRAM resistance for decryption. From references, the minimal energy for a single RRAM programming is approximately $1~\text{pJ}$~\cite{sassine2018sub}, taking about $100.43~\text{ns}$  ~\cite{8942041}, and the required bits to be encrypted is listed in Table~\ref{tab:Modelresult}. At the frequency of $10~\text{MHz}$, we assume RD is capable of simultaneous rewriting of $2^{10}$ RRAMs for the greedy assumption. In contrast, our proposed genetic encryption method (GD) uses the decryptor module for decryption. This requires about $14.75~\text{pJ}$ per decryption per bit, without introducing additional delay due to parallel decryption while computation. However, the decryptor module's limitations restrict the maximum operational frequency to $25~\text{MHz}$. With these parameters, for the five models—NMNIST, DVSGesture, EEGMMIDB, Braille letter, and SHD—we note substantial reductions in decryption energy consumption by factors of $\times1503$, $\times6780$, $\times1497$, $\times474$, and $\times59$ while decryption latency reductions of $\times2050$, $\times4250$, $\times4250$, $\times175$, and $\times500$.

\vspace{-6pt}
\subsection{Security Analysis}
\label{reliability}
\begin{figure}[!t]%
    \centering
    {\includegraphics[width=1\columnwidth]{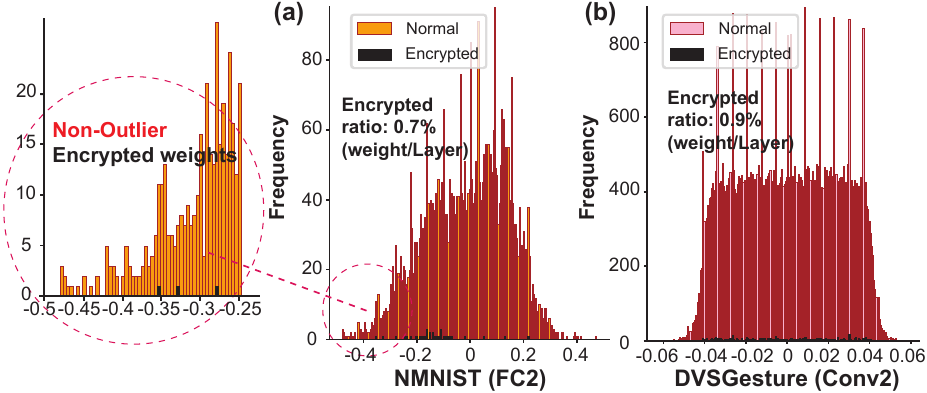}} \\
       \vspace{-10pt}  
    \caption{Stealthy values of synaptic weight distribution for a single SNN encrypted layer: (a) NMNIST FC2 (Acc: 19.7\%) (b) DVSGesture Conv2 (Acc: 9.1\%) from Figure 5.}%
    \label{fig:Distribution_encryptionV3}%
       \vspace{-10pt}  
\end{figure}

\textbf{Stealthiness.} It is legitimate to believe that weights with outlier values may cause more distortion to model outputs and hence be selected. If this holds, attackers may greatly reduce search space by ranking largest values of weights and thus breaking the encryption easily. In Fig.~\ref{fig:Distribution_encryptionV3}, we extract two SNNGX-encrypted models from models encrypted in Fig.~\ref{fig:by_layer}, which are output layer (FC2) encrypted of NMNIST and second convolutional layer (Conv2) encrypted of DVSGesture, to analyze value distribution of encrypted weights. FC2 Layer has weights of size $512 \times 10$ = 5120 and encrypted weights 37. Conv2 Layer has weights of size $64 \times 128 \times 3 \times 3$ = 73728 (a standard VGG layer) and encrypted weights 681. We can observe no matter layer type (Linear vs. Conv layer), layer sequence (Output vs. 2nd Input layer) and datasets and models, SNNGX would not encrypt outlier weights in Fig.~\ref{fig:Distribution_encryptionV3} distribution. They distribute evenly and disperse diffusely across inlier values without a notable pattern, which ensures stealthiness for our encryption method.
\vspace{+2pt}

\begin{figure}[!t]%
    \centering
    {\includegraphics[width=1.0\columnwidth]{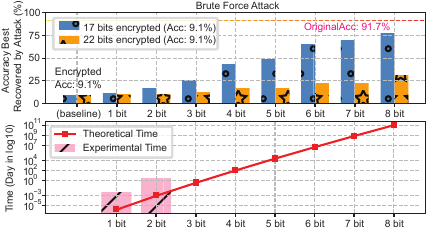}} \\
       \vspace{-10pt}  
    \caption{Brute Force Attack on DVSGesture Conv1 (Acc: 9.1\%), most efficient decryption case of 17 encrypted over 1,152 total weights from Table 3; (Upper): Worst-case from partly recovered key; (Lower): Time cost for Attacking (in Day).}%
    \label{fig:BruteForce_Attack04}%
       \vspace{-14pt}  
\end{figure}

\textbf{Brute Force Attack.} As from Sec. 3.6, the most efficient decryption case is vulnerable to a Brute force attack with $O(\binom{n}{k}\frac{n-k+1}{n-2k+1})$ complexity, as well as encrypted bits black-box vulnerability to a partly recovered key. In Fig.~\ref{fig:BruteForce_Attack04}, we pick and analyse DVSGesture Conv1 from Table 3; this layer has $2 \times 64 \times 3 \times 3$ = 1152 weights and 17 encrypted weights. In Fig.~\ref{fig:BruteForce_Attack04} upper plot, we show that SNNGX encryption is a "group effect". There is no outlier effect for any single bit. Moreover,  a partly recovered key is proven to endanger most efficient decryption cases, 5 bits to recover 40\% accuracy. To demonstrate more bits encrypted increase robustness, we encrypted the same layer an additional time with same protection effect but 22 bits encrypted. We do not follow $\sum^{k}_{i=1}\binom{1152}{i}$ but $\sum^{k}_{i=1}\binom{17}{i}$ and $\sum^{k}_{i=1}\binom{22}{i}$ for k = 1,2...8 bit to compute danger of partly recovered key. Results show encrypting only 5 more bits significantly raises robustness. In Fig.~\ref{fig:BruteForce_Attack04} lower plot, experimental time is measured by actual running of $\sum^{k}_{i=1}\binom{1152}{i}$. The theoretical time is generously estimated by RTX 4090 peak computational capability (82.6 TFLOPs). DVSGesture model's single inference is estimated by \cite{Ao2021EndtoendAD} to be 0.012 TFLOPs. Therefore, maximally $\approx$ 6,900 inferences can be made in one second. This forms the basis for theoretical time estimation. From the plot, 5 bits take 27 years to recover using RTX 4090 on a layer with 1,152 weights (Feasible). 8 bits take 0.27 billion years to recover (Infeasible). We propose the absolute number of weights encrypted \textbf{\textit{should not be less than 30 for any layer}} with reliability.
\begin{figure}[!t]%
    \centering
    {\includegraphics[width=1.\columnwidth]{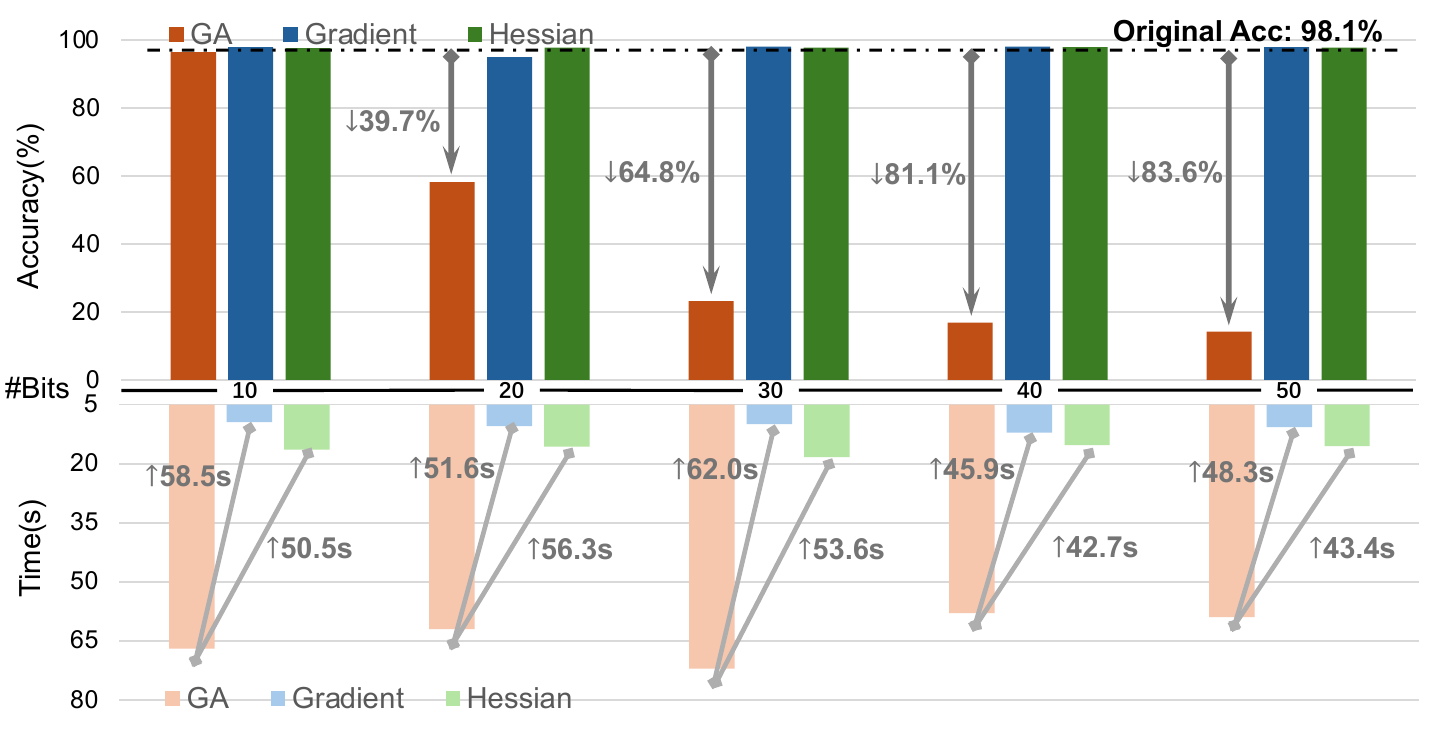}} \\
       \vspace{-10pt}  
    \caption{Comparison of accuracy and efficiency cost of different vulnerability weight identification methods (the final layer of 2312F-512F-10F model ).}%
    \label{fig:hessian}%
       \vspace{-12pt}  
\end{figure}

\vspace{-3pt}
\subsection{Compared to Hessian and Gradient methods} \label{gradient_result}
In this section, we compare the effectiveness and efficiency of the proposed GA method against other Hessian-based \cite{frantar2022optimal} and gradient-based approaches \cite{8942041} for detecting vulnerable weights, which are followed Eq.~\ref{gradient}. As shown in Fig.~\ref{fig:hessian}, using 128 samples for weights of size $512 \times 10$, the GA can significantly reduce the accuracy by encrypting nearly 20 bits. However, for the first-order gradient and Hessian-based methods, even changing 50 bits does not result in a significant change in model accuracy, demonstrating the limitations of gradient methods in SNNs. This further illustrates the effectiveness of our proposed GA method. We also compared the computational time consumption of different methods. Although the GA approach costs more time than the left two methods, it exhibits a clear advantage in terms of encryption effectiveness. The offline time expenditure of several tens of seconds is also acceptable.

\section{Conclusion}
There is a growing interest in protecting the IP of bio-plausible SNNs and making them robust against malicious attacks, similar to how ANNs are protected. In this paper, we introduce SNNGX, which employs a software-hardware co-design approach to encrypt and decrypt bits on RRAM with a genetic search approach, safeguarding the IP of SNNs with minimal hardware cost and great flexibility. We hope that our work will inspire further research in the development of lightweight and effective security algorithms for SNNs.

\section{Acknowledgement}
This research was supported by ACCESS-AI Chip Center for Emerging Smart Systems sponsored by InnoHK funding, Hong Kong SAR, by NSFC (Grant No. 62122004, 62122076), HK RGC (Grant Nos. 27206321, 17205922, 17212923), and by Strategic Priority Research Program of CAS (No. XDB44000000), Key Research Program of Frontier Sciences, CAS (No. ZDBS-LY-JSC012), Youth Innovation Promotion Association CAS.
\vspace{+15pt}

\bibliographystyle{ACM-Reference-Format}
\bibliography{references}

\end{document}